\documentclass[twoside,fleqn]{article}
\usepackage{espcrc2}
\usepackage[dvips]{epsfig}
\title{Dynamics of the 2d Potts model phase transition\thanks{This 
work was in part supported by the US Department of Energy under
contract DE-FG02-97ER41022.} }
\author{Alexander Velytsky\address{Department of Physics, The Florida 
        State University, Tallahassee, FL~32306, USA}$^{,\, \rm c}$,
Bernd A. Berg$^{\rm \,a,}$\address{Okazaki National Research Institute, 
        Okazaki, Aichi 444-8585, Japan}$^{,\, \rm c}$,
Urs M. Heller$^{\rm \,a,}$\address{School of Computational Science and 
Information Technology,\\
  ~~The Florida State University, Tallahassee, FL 32306-4120, USA} }
\begin{document}
\begin{abstract}
The dynamics of 2d Potts models, which are temperature driven through the 
phase transition using updating procedures in the Glauber universality 
class, is investigated.  We present calculations of the hysteresis for 
the (internal) energy and for Fortuin-Kasteleyn clusters. The shape of 
the hysteresis is used to define finite volume estimators of physical
observables, which can be used to study the approach 
to the infinite volume limit. We compare with equilibrium configurations 
and the preliminary indications are that the dynamics leads to considerable 
alterations of the statistical properties of the configurations studied.
\end{abstract}
\date{\today}
\maketitle

\section{Introduction}

Lattice gauge theory investigations of the deconfining phase 
transition, for reviews see~\cite{MO96}, have mainly been limited 
to studies of their equilibrium properties. In contrast to that, 
their occurrence in nature is governed by temperature driven dynamics.
In the early universe it is a cooling process, which is estimated 
to proceed slowly on the scale set by hadronic relaxation times. 
In heavy ion collisions a rapid heating of the nuclei is followed by 
a much slower cooling process and it is no longer clear whether 
this cooling is slow or fast on the relaxation time scale. 
In addition, the small sizes of the nuclei systems impose 
complications which require to be studied.

What dynamics should we study? Time evolution in nature would correspond
to some molecular dynamics simulation of full QCD in Minkowski space, 
which is out of the question as already Euclidean space Monte Carlo (MC) 
calculations of QCD stretch computer resources to their limits. One may, 
however, argue that the computer-time evolution of a MC simulation 
reveals the correct physical features, as long as 
the updating procedure stays in the universality class of Glauber 
dynamics~\cite{Gl63}. These are local MC updating schemes, which imitate 
the thermal fluctuations of nature.
Examples are single- and multi-hit Metropolis algorithms,
as well as the heat-bath algorithm.

Also to simulate the dynamics of the deconfining phase transition 
within Euclidean lattice QCD is out of reach for an exploratory
study. To gain some qualitative understanding, we have decided to 
investigate $q$-state Potts models, 
whose simulations are far less computer time intensive. The 
relationship to QCD is that an effective model for the Polyakov loop 
is provided by the $3d$, 3-state Potts model in an external magnetic 
field~\cite{SY82}. Even this simplification is not yet a 
convenient starting point. To get confidence in our numerical methods,
we limit our first round of simulations to the $q$-state 
Potts models in $2d$, for which a number of rigorous results 
exist~\cite{Ba73,BJ92}.

In the following we drive our systems many times through various 
heating and cooling schedules and measure physical observables along 
the way. Averages and their error bars are calculated with respect to 
the repetitions of the schedule. Our observables are the internal 
energy and properties of Fortuin-Kasteleyn~\cite{FK72} clusters. 


Details of our investigations are reported in the next section and a 
brief summary with conclusions is given in the final section.

\section{Results}

We have performed simulations on $L\times L$ lattices for the 
$2d$ Potts model with $q=2,\,5$ and 10, to have an example of a second 
order transition ($q=2$), a weak ($q=5$) and a strong ($q=10$) first 
order transition. Due to space restrictions we 
limit the following presentation to $q=10$.

For suitably chosen values of $\beta_{\min}$ and $\beta_{\max}$, we 
run in the range $\beta_{\min}\le\beta\le \beta_{\max}$ through each 
hysteresis cycle at least 640 times with $\beta$ changed by 
$\pm\triangle\beta$ after every sweep. The stepsize is
\begin{equation} \label{delta_beta}
\triangle \beta = {2(\beta_{\max}-\beta_{\min}) \over n_{\beta}\, L^2}
\end{equation}
where $\beta_{\min}$ and $\beta_{\max}$ define the terminal temperatures
and the integer $n_{\beta}=1,2,\dots $ is varied. In the limit where 
we send $L^2\to\infty$ first and then $\triangle \beta \to 0$ the 
hysteresis is expected to survive, whereas it disappears if we send 
$\triangle \beta \to 0$ first and then $L^2\to\infty$.

\begin{figure}[ht] \vspace{-2mm} \begin{center}
\epsfig{figure=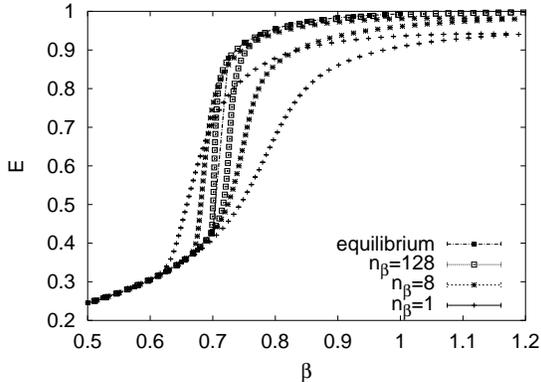,width=\columnwidth} \vspace{-15mm}
\caption{Energy hysteresis curves for the 10-state $d=2$ Potts 
model on a $80\times 80$ lattice for $\beta_{\min}=0.4$,
$\beta_{\max}=1.2$ (not the entire range is shown) and the values 
of $n_{\beta}$ as given in the figure.} 
\label{fig_q10_hys} \end{center} \vspace{-3mm}
\end{figure}

For a first order phase transition, the slowing down of the Markov
process is exponential in computer time, $\sim \exp [2\,f_sL]$,
where $f_s$ is the interfacial tension. In the following we study
the limit $L\to\infty$ and $\triangle \beta (L;n_{\beta})\to 0$ for 
$n_{\beta}$ fixed. For 1-hit Metropolis updating we give in
figure~\ref{fig_q10_hys} our energy hysteresis data for the 
10-state Potts model on a $80\times 80$ lattice. Towards large 
$n_{\beta}$ we see convergence to the equilibrium curve, which is 
obtained by means of a multicanonical~\cite{BN92} simulation. To 
analyze the physical content of such a hysteresis, we define for each 
$n_{\beta}$ finite volume estimators for a physical observable as
its value at the maximum opening of the corresponding hysteresis.

\begin{figure}[ht] \vspace{-2mm} \begin{center}
\epsfig{figure=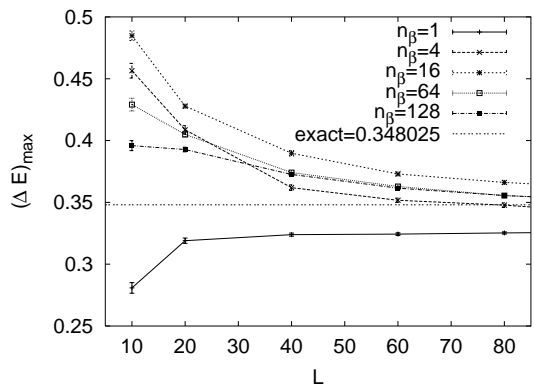,width=\columnwidth} \vspace{-15mm}
\caption{Finite size latent heat estimates for the 10-state $d=2$ 
Potts model versus $L$ for fixed values of $n_{\beta}$, obtained from 
the maximum opening of energy hysteresis curves like those of 
figure~\ref{fig_q10_hys}. The lines are just to guide the eyes. } 
\label{fig_q10_lh} \end{center} \vspace{-3mm} \end{figure}

Figure~\ref{fig_q10_lh} shows, for various fixed values of $n_{\beta}$,
the thus obtained latent heat estimates versus the lattice size $L$. 
For sufficiently large $L$ all the estimators should, independently 
of their $n_{\beta}$ values, converge towards the infinite volume 
value, which is also given in the figure. For each $n_{\beta}$ we 
have performed fits to extrapolate towards $L\to\infty$, but only 
some of them give good estimates. For our small and medium sized 
lattices the results appear to suffer from finite size corrections 
in $L$ as well as in $\triangle\beta$, with the additional twist 
that $\triangle\beta$ can also be too small for some of the lattice 
sizes used. The best $L\to\infty$ estimate is
\begin{equation} \label{lh_q10}
\triangle E = 0.3485\,(12)\ {\rm at}\ n_{\beta}=16\ .
\end{equation}
Results for other values of $q$, other observables and 
details of these fits will be reported elsewhere. 
 

To focus on a detailed physical understanding of the finite volume
transitions, we have begun a cluster analysis of the configurations.
The appropriate cluster definition is the statistical one of Fortuin 
and Kasteleyn~\cite{FK72}, see Ref.\cite{StAh94} for a review of this 
and related topics. We are interested in the effects of 
heating and cooling on the cluster structure, compared to their
equilibrium structure. Whereas nucleation theory holds in the
mean field approximation, phase transitions can also proceed via 
spinodal decomposition as the result of an off-equilibrium quench, 
which may still lead to observable signals, even when there is 
no longer a transition in the strict thermodynamical sense. This 
is somewhat different from the program of Satz~\cite{Sa01}, who 
focuses on geometric clusters and would like to use them as a 
signal when there is no proper transition.

\begin{figure}[ht] \vspace{-2mm} \begin{center}
\epsfig{figure=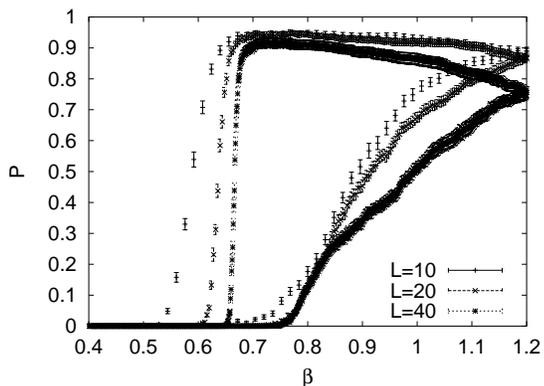,width=\columnwidth} \vspace{-15mm}
\caption{The probability of having a percolating cluster for 
the 10-state $d=2$ Potts model on various lattice sizes 
($n^{hb}_{\beta}=1$). } 
\label{fig_q10_prc} \end{center} \vspace{-3mm} \end{figure}

For the 10-state model we show in figure~\ref{fig_q10_prc} 
our result for percolation of the ordered cluster.
A heat bath algorithm was employed with $n_{\beta}^{hb}=1$.
The figure
provides evidence that the geometry of the Fortuin-Kasteleyn clusters 
is distinct during the cooling and the heating part of the hysteresis. 
The percolation probability of the ordered phase still increases, when 
we already heat the system up again.  A detailed investigation of the
related cluster properties is in progress~\cite{BHMV}.

\section{Summary and Conclusions}

Our energy hysteresis method allows for dynamical estimates of the 
latent heat and works similarly for other physical observables. While 
the precision of these estimates is not competitive with those of
equilibrium investigations~\cite{BN92}, the hysteresis method provides 
information about deviations which are rooted in the dynamics of the 
heating and cooling process. From our ongoing analysis of the 
Fortuin-Kasteleyn clusters we gain insight into the dynamics of the 
transition, which may help to identify signals observable in heavy 
ion experiments.

\end{document}